\documentclass[runningheads]{llncs}
\usepackage{graphicx}
\usepackage{amssymb}
\usepackage{enumitem}
\usepackage{hyperref}
\usepackage{booktabs}
\usepackage{multirow}
\usepackage{lmodern}
\usepackage{listings}
\usepackage{xcolor}
\usepackage{makecell}
\renewcommand{\paragraph}[1]{{\noindent\bfseries #1.}}
\usepackage{environ}
\usepackage{calc}
\newcounter{requirement}
\NewEnviron{requirement}{\refstepcounter{requirement}\fboxsep10pt
\par\medskip%
\noindent%
\fcolorbox{white}{gray!25}{%
    \begin{minipage}{\textwidth-20.8pt}
        \textbf{Req.\,\therequirement:} \BODY
    \end{minipage}}%
\medskip}
\usepackage[left=4.33cm,right=4.33cm,top=4.6cm,bottom=4.7cm]{geometry}
\renewenvironment{abstract}{%
      \list{}{\advance\topsep by0.35cm\relax\small
      \leftmargin=0cm
      \labelwidth=0pt
      \listparindent=0pt
      \itemindent\listparindent
      \rightmargin\leftmargin}\item[\hskip\labelsep
                                    \bfseries\abstractname]}
    {\endlist}
\begin{document}
\title{Natural Language Processing for Requirements Formalization: How to Derive New Approaches?}
%
\titlerunning{Natural Language Processing for Requirements Formalization}
%
\author{Viju Sudhi \and Libin Kutty \and Robin Gröpler}
\authorrunning{Natural Language Processing for Requirements Formalization}
%
\institute{ifak - Institut für Automation und Kommunikation e.V., Magdeburg, Germany\\
\email{vjusudhi@gmail.com}, \email{libinjohn26@gmail.com}, \email{robin.groepler@ifak.eu}}
%
\maketitle              
\begin{abstract}
%
It is a long-standing desire of industry and research to automate the software development and testing process as much as possible.
In this process, requirements engineering (RE) plays a fundamental role for all other steps that build on it.
Model-based design and testing methods have been developed to handle the growing complexity and variability of software systems.
However, major effort is still required to create specification models from a large set of functional requirements provided in natural language.
Numerous approaches based on natural language processing (NLP) have been proposed in the literature to generate requirements models using mainly syntactic properties. Recent advances in NLP show that semantic quantities can also be identified and used to provide better assistance in the requirements formalization process.
In this work, we present and discuss principal ideas and state-of-the-art methodologies from the field of NLP in order to guide the readers on how to create a set of rules and methods for the semi-automated formalization of requirements according to their specific use case and needs.
We discuss two different approaches in detail and highlight the iterative development of rule sets. The requirements models are represented in a human- and machine-readable format in the form of pseudocode.
The presented methods are demonstrated on two industrial use cases from the automotive and railway domains.
It shows that using current pre-trained NLP models requires less effort to create a set of rules and can be easily adapted to specific use cases and domains.
In addition, findings and shortcomings of this research area are highlighted and an outlook on possible future developments is given.
\makeatletter{\renewcommand\@makefntext[1]%
{\noindent\makebox[0pt][r]{}#1}
\footnotetext{\scriptsize This is a preprint of the following chapter: Viju Sudhi, Libin Kutty, and Robin Gröpler, Natural Language Processing for Requirements Formalization: How to Derive New Approaches?, published in Concurrency, Specification and Programming: Revised Selected Papers from the 29th International Workshop on Concurrency, Specification and Programming (CS\&P’21), Berlin, Germany, edited by B.\ Schlingloff et al., 2023, Springer reproduced with permission of Springer Nature Switzerland AG. The final authenticated version is available online at: \url{https://doi.org/10.1007/978-3-031-26651-5_1}. \\
A detailed implementation and data can be found at: \url{https://github.com/ifak-prototypes/nlp_reform}. \vspace{-12pt}
}\makeatother}

\end{abstract}

\section{Introduction}
\label{sec:introduction}

\begin{figure}[tb]
\centering
\includegraphics[width=0.66\linewidth]{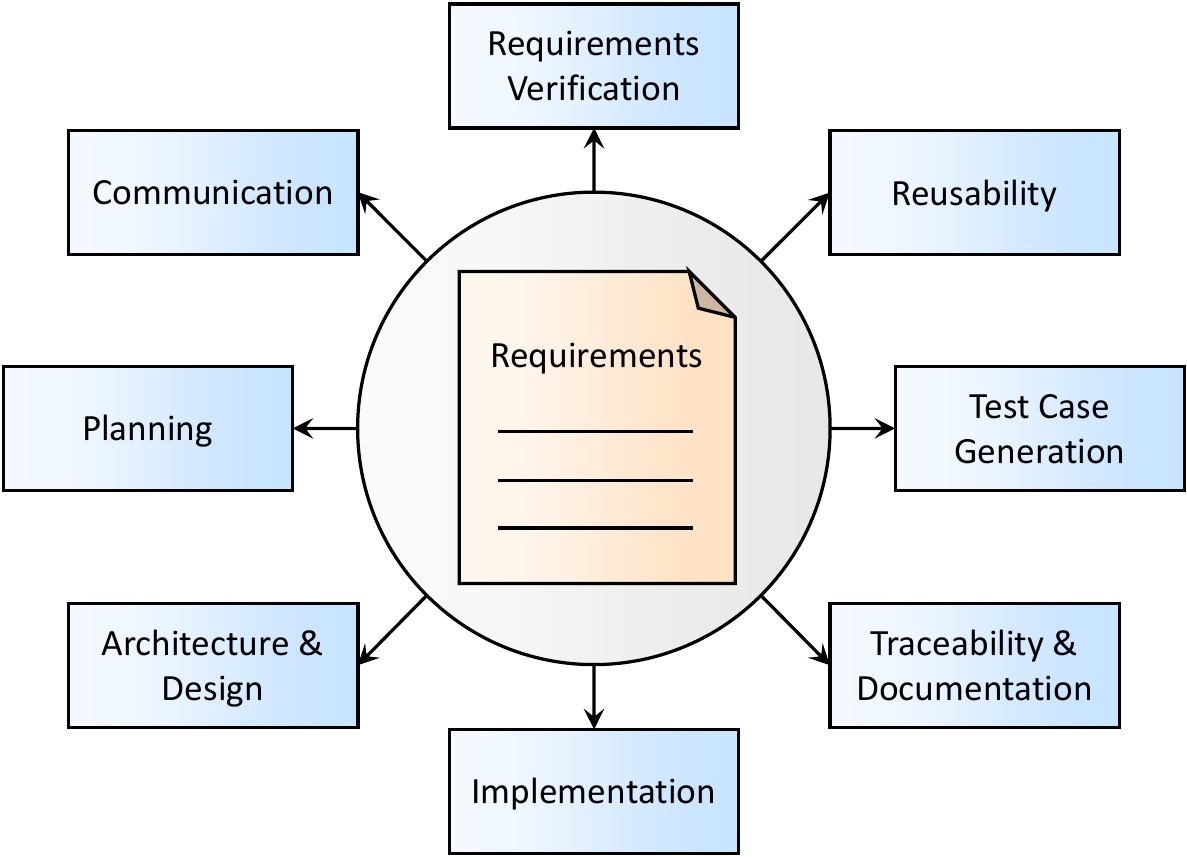}
\caption{Requirements as the basis for the software development and testing process.}
\label{fig:req_central}
\end{figure}

Requirements engineering (RE) plays a fundamental role for the software development and testing process. There are usually many people involved in this process, such as the customer or sponsor, the users from different areas of expertise, the development and testing team, and those responsible for the system architecture. Therefore, requirements are intentionally written in a textual form in order to be understandable for all stakeholders of the software product to be developed.
However, this also means that requirements have an inherently informal character due to the ambiguity and diversity of human language. This makes it difficult to automatically analyze the requirements for further processing. There are many different processes that build on them, such as requirements verification and test case generation (see Fig.~\ref{fig:req_central}).


In order to handle the growing complexity and variability of software systems, model-based design and testing methods have been developed \cite{ammann2017introduction}. 
Especially for safety-critical systems such as in the automotive, railway and aerospace domains, extensive testing based on the requirements is necessary.
However, the manual creation of requirements models from natural language requirements is time-consuming and error-prone, and also requires a lot of expert knowledge.
This is especially true in agile software development which involves continuous improvements and many changes to requirements.

Numerous approaches based on natural language processing (NLP) have been proposed in the literature to generate requirements models using mainly syntactic properties \cite{zhao2021natural,koscinski2021natural}.
Recent advances in NLP show that semantic quantities can also be identified and used to provide better assistance in the formalization of unrestricted natural language requirements \cite{nayak2022req2spec,fritz2021automatic}.
However, most studies propose concrete solutions that work well only for a very specific environment.

The aim of this work is to focus on principal ideas and state-of-the-art methodologies from the field of NLP to automatically generate requirements models from natural language requirements.
We iteratively derive a set of rules based on NLP information in order to guide readers on how to create their own set of rules and methods according to their specific use cases and needs.
In particular, we want to investigate the question: \textit{How to derive new approaches for requirements formalization using natural language processing?} We highlight and discuss the necessary stages of an NLP-based pipeline towards the generation of requirements models.
We present and discuss two approaches in detail: (i) \textit{a dependency and part-of-speech-based approach}, and (ii) \textit{a semantic-role-based approach}. This significantly extends and enhances our previous work \cite{gropler2021nlp}.
The approaches are demonstrated on two industrial use cases: a battery charging approval system from the automotive domain and a propulsion control system in the railway domain.
The requirements models are represented in a human- and machine-readable format in a form of pseudocode \cite{gropler2022automated}.
In summary, this work aims to provide more general and long-lasting instructions on how to develop new approaches for NLP-based requirements formalization.

In Section \ref{sec:methodology}, we present our proposed NLP-based pipeline, introduce the use cases, and present the two different approaches for NLP-based requirements formalization.
In Section \ref{sec:related_work}, we review the state of the literature on the various NLP methods for requirements formalization. 
In Section \ref{sec:discussion}, we discuss some general findings and shortcomings in this area of research and provide an outlook on possible future developments. 
Finally, Section \ref{sec:conclusion} concludes our work.

\section{Methodology}
\label{sec:methodology}

In this section, we propose an NLP-based pipeline for requirements formalization. We first give a brief overview of the use cases and then present and discuss the two different approaches in detail and demonstrate the iterative development of rule sets.

\begin{figure}[tb]
\centering
\includegraphics[width=\linewidth]{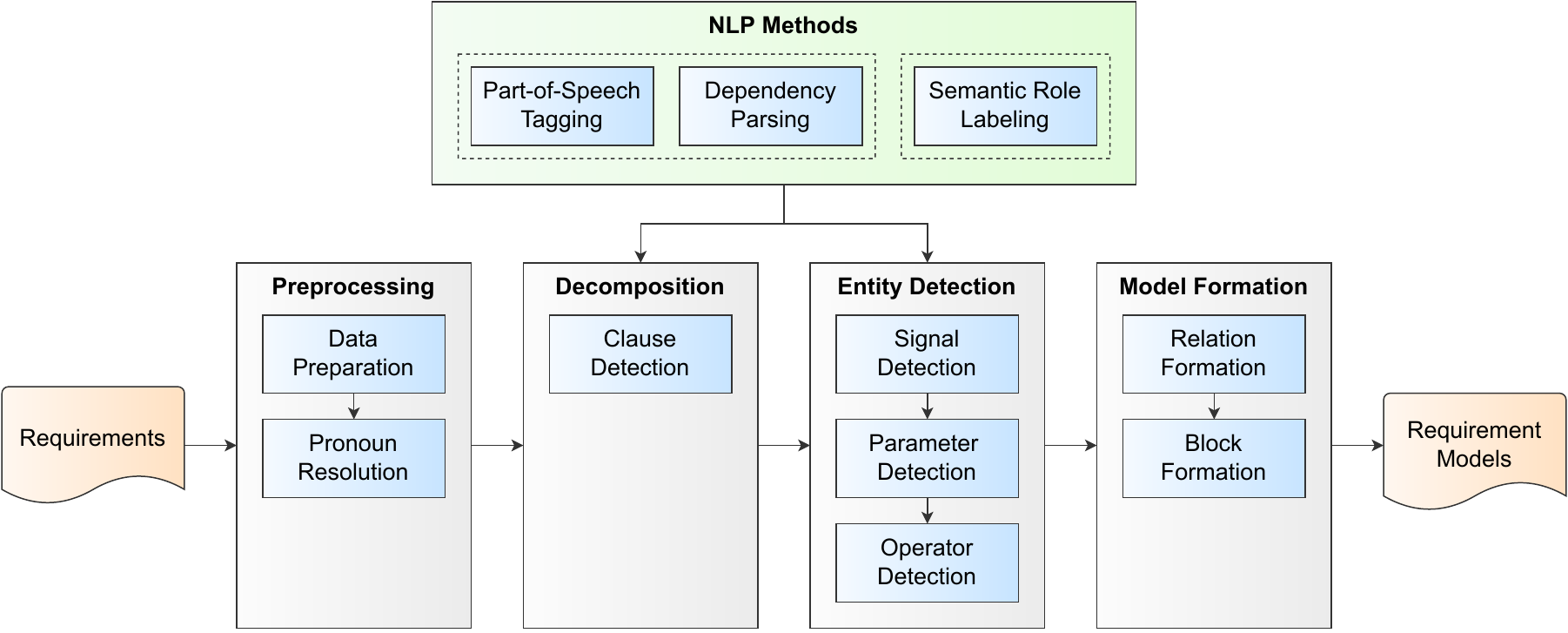}
\caption{Proposed pipeline for requirements formalization.}
\label{fig:pipeline}
\end{figure}

The automatic generation of requirements models from functional requirements written in unrestricted natural language is highly complex and needs to be divided into several steps. 
Therefore, our pipeline consists of several stages, as illustrated in Figure \ref{fig:pipeline}. The different stages of the pipeline are described in more detail below. \vspace{10pt}

\paragraph{Stage 1: Preprocessing} We start with preprocessing the requirements. According to our use cases, we perform data preparation, where we clean up the raw text, and resolve pronouns. This stage need not be limited to these steps and should be flexibly adapted to the style and domain of the requirements at hand. \vspace{10pt}

\paragraph{Stage 2: Decomposition} In order to extract individual actions, the preprocessed requirements are decomposed into clauses. Industrial requirements tend to be complex in certain cases with multiple sentences combined together to convey a particular system behavior. 
We decompose each requirement sentence at certain conjunctions and linking words (\textit{if}, \textit{while}, \textit{until}, \textit{and}, \textit{or}, etc.), assuming that each clause contains a single action. Multi-line and multi-paragraph requirements should also be decomposed into clauses. \vspace{10pt}

\paragraph{Stage 3: Entity detection} In this stage, we use the syntactic and semantic information of each clause to identify the desired model entities, such as \textit{signals}, \textit{components}, and \textit{parameters}. We construct a rule-based mapping that is iteratively derived from the considered requirements of our specific use cases. Further, we map comparison words to \textit{operator} symbols ($<$, $>$, $==$, etc.) using a dictionary. These rules can be easily adapted to different use cases and domains by following the same procedure as described below. \vspace{10pt}

\paragraph{Stage 4: Model formation} 
In the final stage of the pipeline, we assemble the retrieved information to form \textit{relations} for each clause. These can either be assignments of the form \verb|signal(parameter*)| with optional parameter, or conditional statements of the form \verb|signal() operator parameter|. Then we combine them according to the derived logical operators, and assign them to specific \textit{blocks} (\textit{if}, \textit{then}, \textit{else}, etc.). This yields a requirements model for each requirement.

The whole pipeline can be tailored according to the use case and the desired output.
For example, no components (actors) are explicitly mentioned in the requirements we consider.
Therefore, we have omitted this entity and assume that this information is given.
The decomposition and entity detection stages are particularly based on information provided by NLP methods. We present two approaches to handle these stages: (i) \textit{a dependency- and part-of-speech-based approach} (Section \ref{sec:dep-pos}), which utilizes the grammatical constructs of the requirement text, and (ii) \textit{a semantic-role-based approach} (Section \ref{sec:srl-based}), which makes use of the semantic roles given by a pre-trained model. Both approaches work with unrestricted natural language and can be further refined and adapted to different styles and domains according to the needs of the use case.

\subsection{Use cases}
We demonstrate the derivation of our approaches using functional requirements from two different industrial use cases. The first use case is a battery charging approval system provided by AKKA from the automotive domain. The use case describes a system for charging approval of an electric vehicle in interaction with a charging station. In total, AKKA has provided 14 separate requirement statements. 
The requirements are used for a model-based software development and implementation of the functionality in an electronic control unit (ECU).
More details about the industrial use case can be found in \cite{grujic2021testing,gropler2021nlp}.

The second industrial use case is a propulsion control system (PPC) provided by Alstom from the railway domain.
The PPC is part of a large, complex, safety-critical system. It handles the control of the entire propulsion system, including both control software as well as the electrical functions.
Alstom has provided 31 requirements which do not follow a prescribed format in order not to focus on syntax when writing them.
The system architecture and software modules are modeled in Matlab Simulink using a model-based design approach. More information about this use case is given in \cite{simonson2018mbd,gropler2022automated}. 

Note that we show the requirements in a generalized form, as the data we use is confidential.
For demonstration purposes, we have also partially modified the original requirements.

\subsection{Dependency- and Part-of-Speech-Based Approach}
\label{sec:dep-pos}

One possible approach to arrive at formal requirements models is to investigate the grammar of the natural language requirements. For instance, one can make use of dependency tags, part-of-speech tags or combine both of these syntactical information to arrive at the entities for the desired requirements models.

\begin{figure}
\centering
\includegraphics[width=0.95\linewidth]{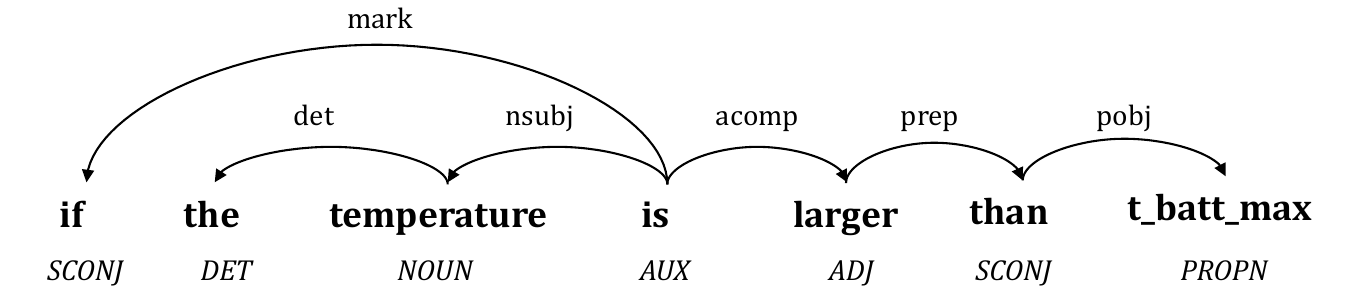}
\caption{Dependency and POS tags for an exemplary requirement phrase.}
\label{fig:dep_pos_ex}
\end{figure}

For example, consider the requirement phrase \textit{``if the temperature is larger than t\_batt\_max''}.
When we parse the dependency and POS tags of this phrase with the help of toolkits like spaCy\footnote{\url{https://spacy.io/}}\footnote{\url{https://explosion.ai/demos/displacy}}, we arrive at a directed graph as shown in Fig.~\ref{fig:dep_pos_ex}. It illustrates that the \textit{root} verb of the phrase is ``\textit{is}" (from which the graph emerges) and shows how each word is associated with this \textit{root} verb. These associations or dependencies are represented by different tags. For example, ``\textit{temperature}" is the nominal subject (\textit{nsubj}) of the \textit{root}, ``\textit{t\_batt\_max}" is the prepositional object (\textit{pobj}) of the \textit{root}, etc. Similarly, we can find the POS tag of each word in the graph, e.g.\ ``\textit{temperature}" is a \textit{noun}, ``\textit{larger}" is an \textit{adjective}, etc.
Some dependency and POS tags are presented in Table \ref{tab:dep_pos_list}.
We suggest the reader to gather an overview of the dependency and POS tags from the generic framework presented as universal dependencies \footnote{\url{https://universaldependencies.org/u/dep/index.html}}\footnote{\url{https://universaldependencies.org/u/pos/index.html}}.

\begin{table}[tb]
    \centering
    \caption{A few commonly used dependency and POS tags}
    \label{tab:dep_pos_list}
    \begin{tabular}{ l@{\extracolsep{10pt}}l }
    \toprule
    \textbf{Dependency tag} & \textbf{Description} \\
    \midrule
    nsubj & Nominal subject - does the action \\
    root & Root of the sentence - the action \\
    dobj & Direct object - on which the action is done \\
    mark & Marker - marks a clause as subordinate to another \\
    conj & Conjunct - indicates a coordinating conjunction \\
    \midrule
    \textbf{Part-of-speech tag} & \textbf{Description} \\
    \midrule
    NOUN & Nouns denote a person, place, thing, animal or idea \\
    ADJ & Adjectives modify nouns \\
    ADV & Adverbs modify verbs \\
    DET & Determiners indicate the reference of the nouns \\
    SCONJ & Subordinating conjunction \\
    \bottomrule
    \end{tabular}
\end{table}

With this background on dependencies and POS tags, we further discuss how we tailor our pipeline with a rule base depending on the syntactic structure of the requirement text. 
For better comprehensibility for readers, the requirements are presented in the order of growing complexity.

\begin{requirement} \label{req:akka_1}
The error state is `E\_BROKEN', if the temperature of the battery is larger than t\_batt\_max.
\end{requirement}

As an entry point to the discussion, consider a simple requirement, Req.\,\ref{req:akka_1}.
Here, the requirement has two primitive actions - one that talks about the ``\textit{error state}" and another that talks about the condition when this ``\textit{error state}" occurs. Similar to this requirement, individual industrial requirements are often composed of multiple actions, making them inherently difficult to process with information extraction toolkits. Hence, as shown in Fig.~\ref{fig:pipeline}, we propose a decomposition step to initially decompose a long complex requirement into clauses which individually explain primitive actions. \vspace{4pt}

\paragraph{Decomposition (conditions)}
To decompose this requirement to individual clauses, we can use the conditional keyword ``\textit{if}" as the boundary. This yields us the following clauses: \vspace{-6pt}
\begin{itemize}
    \item The error state is `E\_BROKEN'
    \item if the temperature of the battery is larger than t\_batt\_max
\end{itemize} \vspace{10pt}

\paragraph{Entity detection} 
For the first clause, we extract the subject of the clause as \textit{``error state''} and the object as \textit{``E\_BROKEN''}. The root verb in the clause is \textit{``is''} which decides how to form the relation. By mapping the subject (with an inflection of the root verb) to the \textit{signal} and the object as the \textit{parameter}, we end up with the relation \verb|set_error_state(E_BROKEN)|. In the second clause, with similar rules, we extract the subject of the clause as \textit{``temperature of the battery''} and the object as \textit{``t\_batt\_max''}. \vspace{10pt}

\paragraph{Operator detection}
However, the root verb occurs in conjunction with a comparative term \textit{``larger"}. The appropriate operator for this word can be fetched with the help of a hyperlinked thesaurus like Roget's Thesaurus\footnote{\url{https://sites.astro.caltech.edu/~pls/roget/}}. This yields us the symbol ``$>$'' indicating the quality ``greatness'', see Table \ref{tab:comp_symbols}. The occurrence of a comparative term differentiates how we form the relation for this clause from the previous clause. The relation formed  from this clause will be: \verb|temperature_of_battery() > t_batt_max|.

\begin{table}[tb]
    \caption{Detection of comparison operators using Roget's Thesaurus}
    \label{tab:comp_symbols}
    \centering
    \begin{tabular}{ l@{\extracolsep{10pt}}l@{\extracolsep{10pt}}c } 
    \toprule
    \textbf{Quantity} & \textbf{Words} & \textbf{Operator} \\
    \midrule
    Superiority & exceed, pass, larger, greater, over, above, ... & $>$ \\ 
    
    Greatness & excessive, high, extensive, big, enlarge, ... & $>$ \\
    \midrule
    Inferiority & smaller, less, not pass, minor, be inferior, ... & $<$ \\
    Smallness & below, decrease, limited, at most, no more than, ... & $<$ \\
    \midrule
    Sameness & equal, match, reach, come to, amount to, ... & $==$ \\
    \bottomrule
    \end{tabular}
\end{table}

\begin{requirement} \label{req:akka_2}
The error state is `E\_BROKEN', if the temperature of the battery is larger than t\_batt\_max and \underline{it} is smaller than t\_max.
\end{requirement}

Now consider a slightly different requirement with a different formulation in the second clause as shown in Req.~\ref{req:akka_2}. \vspace{10pt}

\paragraph{Decomposition (root conjunctions)}
In this clause, the conjunction \textit{``and''} occurs. Unless we decompose this conjunction, the clause by itself does not address a single primitive action. In this case, the conjunction \textit{``and''} connects the two root verbs \textit{``is''} (before \textit{``larger''}) and \textit{``is''} (before \textit{``smaller''}). Such root conjunctions can be decomposed by considering the conjunction as the boundary. This yields us the following clauses:
\begin{itemize}
    \item if the temperature of the battery is larger than t\_batt\_max
    \item \underline{it} is smaller than t\_max
\end{itemize}

\paragraph{Pronoun resolution}
The first clause is similar to the one presented in the previous requirement. However, the second clause presents a new challenge. The pronoun \textit{``it''} needs to be resolved before proceeding to entity detection. We propose a simple pronoun resolution step to find the third person pronouns (singular: \textit{it}, plural: \textit{they}) by replacing each occurrence of a pronoun with the farthest subject. In this clause, we replace \textit{``it''} with \textit{``the temperature of the battery''}. This yields the clause \textit{``if the temperature of the battery is smaller than t\_max''}. This is again similar to the discussed clauses and is handled with the same rules. 

A further check on the grammatical number of the pronoun and its antecedent is advised, if the requirement quality is in doubt. The pronouns without an antecedent (called pleonastic pronouns) should not be resolved. We also assume first person or second person pronouns hardly occur in industrial requirements.

\begin{requirement} \label{req:akka_3}
The error state is `E\_BROKEN', if the temperature of the battery is larger than t\_batt\_max \underline{and} t\_max.
\end{requirement}

Now consider a requirement which has the following clause as given in Req.\,\ref{req:akka_3}. Unlike the clause in the previous requirement, this clause has a conjunction \textit{``and''} between two noun phrases \textit{``t\_batt\_max''} and \textit{``t\_max''}. \vspace{10pt}

\paragraph{Decomposition (noun phrases)}
This demands a slightly more complex noun phrase decomposition. Unlike the other decomposition steps described above, this step further requires to extract the subject phrase of the first noun phrase and prefix it to the latter. This yields us the following clauses:
\begin{itemize}
    \item if the temperature of the battery is larger than t\_batt\_max
    \item if the temperature of the battery is larger than t\_max
\end{itemize}
These clauses are identical to the discussed clauses and hence, the entities are extracted with the same rules.

\begin{requirement} \label{req:akka_4}
The error state is `E\_BROKEN', if the temperature of the battery is \underline{between} t\_batt\_max and t\_max.
\end{requirement}

Another clause is shown in Req.\,\ref{req:akka_4}, with a connector ``\textit{between}" in the text. \vspace{10pt}

\paragraph{Decomposition (connectors like ``between")}
To decompose such clauses, we can replace \textit{between A and B} with the construct \textit{greater than A and less than B}. This however, assumes the real values of \textit{A} to be less than \textit{B}. Although the assumption is true in most cases, we advise to further validate the decomposition, e.g.\ by integrating a user feedback loop. With the above assumption, the decomposition of this clause results in the following clauses:
\begin{itemize}
    \item if the temperature of the battery is greater than t\_batt\_max
    \item if the temperature of the battery is less than t\_max
\end{itemize}


\begin{requirement} \label{req:akka_5}
The charging approval shall be given if the connection with the charging station is \underline{active}.
\end{requirement} \vspace*{-6pt}

\begin{requirement} \label{req:akka_6}
The charging approval shall \underline{not} be given if the connection with the charging station is \underline{inactive}.
\end{requirement}

Requirements can also contain negations. Consider the pair of requirements Req.\,\ref{req:akka_5} and Req.\,\ref{req:akka_6}, which when decomposed yield the following clauses:

\begin{itemize}
    \item \textit{R5C1:} The charging approval shall be given
    \item \textit{R5C2:} if the connection with the charging station is active
\end{itemize}

\begin{itemize}
    \item \textit{R6C1:} The charging approval shall not be given
    \item \textit{R6C2:} if the connection with the charging station is inactive
\end{itemize}

\paragraph{Entity detection (negation handling)}
The corresponding clauses of these requirements contradict each other, i.e. \textit{R5C1} and \textit{R6C1} contradict each other as well as \textit{R5C2} and \textit{R6C2}. In \textit{R5C1}, there is no explicit negation. Hence, we can handle this clause just as any other clause explained before. However, the root verb in \textit{R6C1} \textit{``be''} occurs in conjunction with a \textit{``not''}. This can be addressed by introducing a negation operator (a logical \verb|not|) in the relation. For example, \textit{R5C1} yields \verb|give_charging_approval()| and \textit{R6C1} yields \verb|not give_charging_approval()|.



Handling the second clauses of the requirements poses a different challenge. In \textit{R5C2}, the word \textit{active} occurs, while in \textit{R6C2}, the antonym of this word \textit{inactive} occurs. We propose to assume the first occurrence of a word qualifying such boolean expressions as \textit{boolean true} and if its antonym is cited in a different requirement, it can be assigned as \textit{boolean false}. The antonyms can be identified using a hierarchical thesaurus like WordNet\footnote{\url{http://wordnetweb.princeton.edu/perl/webwn}}. We assume this way of approaching negations is more intuitive than looking at a single word and inferring its sentiment. The sentiment of a word does not necessarily help us define the boolean value of the word.


In the light of the above discussion, we present an overview of a few rules we use to extract the entities in the following tables.
The dependency and POS tags are first mapped to the syntactic entities \textit{subject}, \textit{object} and \textit{predicate}, see Table \ref{tab:dep_pos_constructs1}. Then these syntactic entities are mapped to the model entities \textit{signal} and \textit{parameter} according to specific rules, see Table \ref{tab:dep_pos_constructs2}. The \textit{operator} is identified from the ADJ denoted by op(ADJ) in the table.
The rules can be extended e.g.\ by considering the type of action (nominal, boolean, simple, etc.) and verb types (transitive, dative, prepositional, etc.). \vspace{10pt}

\begin{table}
\caption{Mapping of DEP/POS tags to syntactic entities}
\label{tab:dep_pos_constructs1}
\centering
\begin{tabular}{ l@{\extracolsep{20pt}}l }
\toprule
\textbf{DEP/POS tags (constituents)} & \textbf{Syntactic entities} \\
\midrule
nsubj, nsubjpass & subject \\
pobj, dobj & object \\
root, root + NOUN, & \multirow{2}{*}{predicate} \\
auxpass/root + ADP/ADV/ADJ \\
\bottomrule
\end{tabular}
\end{table}


\begin{table}
    \caption{Mapping of syntactic entities to model entities}
    \label{tab:dep_pos_constructs2}
    \centering
    \begin{tabular}{ l@{\extracolsep{16pt}}l@{\extracolsep{10pt}}l@{\extracolsep{10pt}}l }
    \toprule
    \textbf{Syntactic entities} & \textbf{Signal} & \textbf{Operator} & \textbf{Parameter}\\
    \midrule
    subject, predicate & predicate\_subject & - & - \\
    subject, predicate, object & predicate\_subject & - & object \\
    ADJ, subject, predicate, object & predicate\_ADJ & - & object \\
    \midrule
    subject, ADJ, object & subject & op(ADJ) & object \\
    \bottomrule
    \end{tabular}
\end{table}

\paragraph{Discussion}
Although a custom rule base exploiting dependency and POS tags is well suited for a particular use case, it demands a tremendous effort to generalize the rules for requirements across domains varying in nature and style. With each new style of requirement, the rule base may need to be updated, leading to further effort and deliberation. We argue this by considering the number of dependency and POS tags, which are 37 dependency tags and 17 POS tags according to the revised universal dependencies \cite{marneffe2021universal}. 

\subsection{Semantic-Role-Based Approach}
\label{sec:srl-based}

We can also formalize requirements by extracting the semantic roles in the requirement text. The semantic roles represent the relation between the words or phrase in regards to the main verb present in the sentence. It describes the conceptual meaning behind the word.

\begin{figure}
\centering
\includegraphics[width=0.4\linewidth]{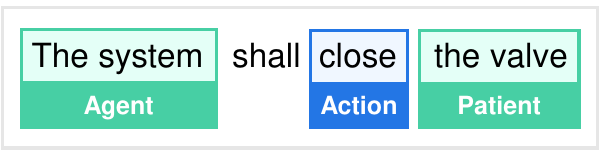}
\caption{Semantic roles for an exemplary requirement phrase.}
\label{fig:srl_frames_ex}
\end{figure}

For example, consider a simple requirement phrase, ``\textit{The system shall close the valve}". As illustrated in Fig.\ \ref{fig:srl_frames_ex}, \textit{``the system''} is semantically the \textit{agent} of the action \textit{``close''}. Further, this action is performed on \textit{``the valve''} which is semantically its \textit{patient}. In general, the \textit{agent} is the one who initiates the action and the \textit{patient} is the one on whom the action is performed. These roles are semantic properties in contrast to \textit{subject} and \textit{object}, we used in Section \ref{sec:dep-pos}, which are syntactic properties. 
Some semantic roles are presented in Table \ref{tab:semantic_roles}.

\begin{table}
\caption{List of basic semantic roles}
\label{tab:semantic_roles}
\centering
\begin{tabular}{ l@{\extracolsep{20pt}}l }
 \toprule
 \textbf{Role} & \textbf{Description} \\
 \midrule
 Agent & Who/what performed the action \\ 
 Patient & Who/what was the action performed on \\
 Location & Where the action was performed\\
 Time & When the action was performed \\
 Instrument & What was used to perform the action \\
 Beneficiary & Who/what is benefited by the action \\
 \bottomrule
\end{tabular}
\end{table}

Semantic Role Labeling (SRL) is the task of assigning these roles to the phrases or words that describe their semantic meaning. SRL is one of the leading tasks in natural language processing \cite{marquez2008semantic}. It can be used for applications such as question answering, data mining etc. We use SRL for requirements formalization by identifying the roles to form the desired requirements models. We use a state-of-the-art pre-trained deep learning BERT model \cite{shi2019simple} by AllenNLP\footnote{\url{https://demo.allennlp.org/semantic-role-labeling}}. This model generates frames, each containing a set of arguments related to a verb in the sentence. 
The arguments are defined based on English Propbank defined by Bonial et al.\ \cite{bonial2014propbank}. Each argument describes the semantic relation of a word to the verb with which it is associated. The arguments can be either numbered or functional. Table \ref{tab:srl_arguments} shows some of the arguments and the corresponding roles.

\begin{table}[ht]
\caption{Basic arguments based on English Propbank}
\label{tab:srl_arguments}
\centering
\begin{tabular}{ l@{\extracolsep{20pt}}l@{\extracolsep{20pt}}l }
 \toprule
  & \textbf{Argument} & \textbf{Role} \\
 \midrule
 \multirow{5}{*}{Numbered} & ARG0 & Agent \\
  & ARG1 & Patient \\
  & ARG2 & Instrument/beneficiary \\
  & ARG3 & Start point/beneficiary\\
  & ARG4 & End point \\
\midrule
 \multirow{4}{*}{Functional} & ARGM-LOC & Location \\
  & ARGM-TMP & Time \\
  & ARGM-NEG & Negation \\
  & ARGM-PRD & State \\
 \bottomrule
\end{tabular}
\end{table}

We will now demonstrate how we handle the decomposition and entity detection stages with the SRL roles. This can help in considerably reducing the number of rules compared to the first approach while handling even more complex requirements. It is worth to note that there are only 24 SRL arguments (5 numbered and 19 functional) and we argue that even an exhaustive rule base would only demand a combination of some of these different arguments.

\begin{requirement} \label{req:bomb_1}
The maximum power shall be limited to [G\_Max] and the event ``High device temperature" shall be indicated when the device temperature exceeds [T\_Hi] ºC.
\end{requirement}

Consider the requirement Req.\,\ref{req:bomb_1}. This requirement has three desired clauses with two statements based on one condition. \vspace{10pt}

\paragraph{Preprocessing}
In the requirement, we have the name of an event \textit{``High device temperature''} within quotes which describes what happens when this event occurs in the system. In certain other cases, the use case has longer event names which by itself had semantic roles within. To help the SRL model to distinguish the event names from the rest of the requirement text, we preprocess the requirement by replacing the occurrences of event names in quotes with abbreviated notations like \textit{``E1''}. We also eliminate square brackets and units around variables, for example \textit{``G\_Max''} and \textit{``T\_Hi''} in Req.\,\ref{req:bomb_1}.

\begin{figure}
\centering
\includegraphics[width=0.75\linewidth]{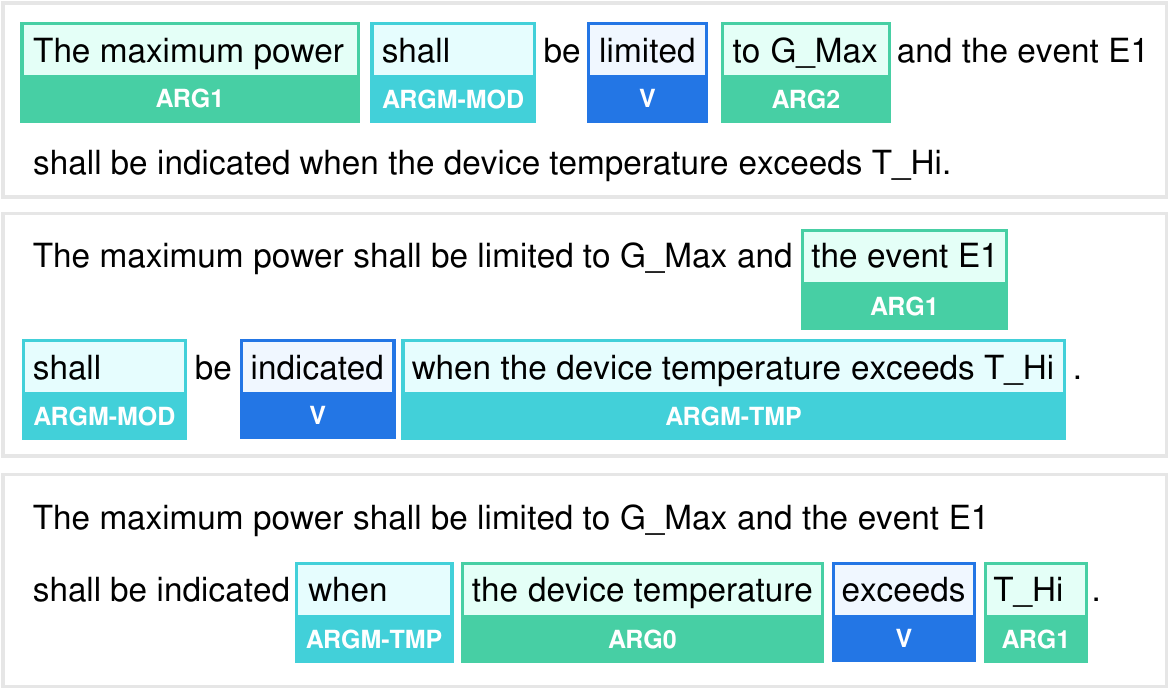}
\caption{SRL frames for Req.\,\ref{req:bomb_1}.}
\label{fig:srl_frames1}
\end{figure}

The pre-trained SRL model also retrieves frames for modal verbs like \textit{``be''} or \textit{``shall''} which may not have any related arguments. To devise a rule set that works with almost all frames, we discard frames with just one argument. \vspace{10pt}

\paragraph{Decomposition} 
Unlike the decomposition described in Section. \ref{sec:dep-pos} which utilized dependency and POS tags, here we discuss the decomposition of complex requirements based on the detected SRL frames. As shown in Fig. \ref{fig:srl_frames1}, we obtain a total of three frames from the pre-trained SRL model applied on Req. \ref{req:bomb_1}. The span (obtained by adding up the requirement phrases belonging to each role in the frame) of these detected frames yield us the following decomposed clauses:

\begin{itemize}
    \item the maximum power shall limited to G\_Max
    \item the event E1 shall indicated
    \item when the device temperature exceeds T\_Hi
\end{itemize}

In the second clause, we have the role \textit{ARGM-TMP} which describes the condition of the action. Here, we have \textit{ARGM-TMP} with the desired conditional clause as its span which tells us that this condition is related to this particular action. But when we look at the requirement we know that the condition clause applies to both the other clauses. So as to avoid wrong formation of the output model, we will avoid those \textit{ARGM-TMP} role with the full clause as span and only consider those \textit{ARGM-TMP} roles with a single word like \textit{``when''} or with the unit of time. \vspace{10pt}


\paragraph{Entity detection}
The first frame is associated with the verb \textit{``limited''}. We have \textit{``The maximum power"} as \textit{ARG1} (describing on whom the action is performed on), \textit{``to G\_Max"} as \textit{ARG2} (an instrument). We can map the verb together with \textit{ARG1} as the signal and \textit{ARG2} as the parameter. Note that we use the lemma (basic form) of the verb to form the signal. Also, the stop words in the arguments are eliminated. This yields us the relation for this clause as: \verb|limit_maximum_power(G_Max)|. From this relation, we form the construct \verb|V_ARG1(ARG2)| which can be further used if we get similar frames.

The second frame, after avoiding the argument \textit{ARGM-TMP}, follows a similar construct as the first frame only without the optional parameter. We also perform back mapping of the abbreviated event name \textit{``E1''} with its actual event name yielding the relation: \verb|indicate_event_high_device_temperature()|.

The third frame is a conditional clause which is identified by the argument \textit{ARGM-TMP} with a span of \textit{``when"}. Here, we have \textit{``the device temperature"} as \textit{ARG0} (the agent) and \textit{``T\_Hi"} as \textit{ARG1} (the patient). We map \textit{ARG0} as the signal and \textit{ARG1} as the parameter. \vspace{10pt}

\paragraph{Operator detection}
Additionally, a comparison word \textit{``exceeds''} occurs in the role \textit{V} of this clause. We detect the corresponding operator by looking it up in Roget's Thesaurus, similar to the first approach. In this case, the word \textit{``exceeds"} gives the symbol ``$>$''. Thus, this clause translates to the relation \verb|device_temperature() > T_Hi|. We build a general construct for similar frames with operator as \verb|ARG0 V ARG1|.

\begin{requirement} \label{req:bomb_2}
The device fuel pump shall \underline{not} be activated until the fuel level falls below [L\_Fp].
\end{requirement}

\begin{figure}
\centering
\includegraphics[width=0.96\linewidth]{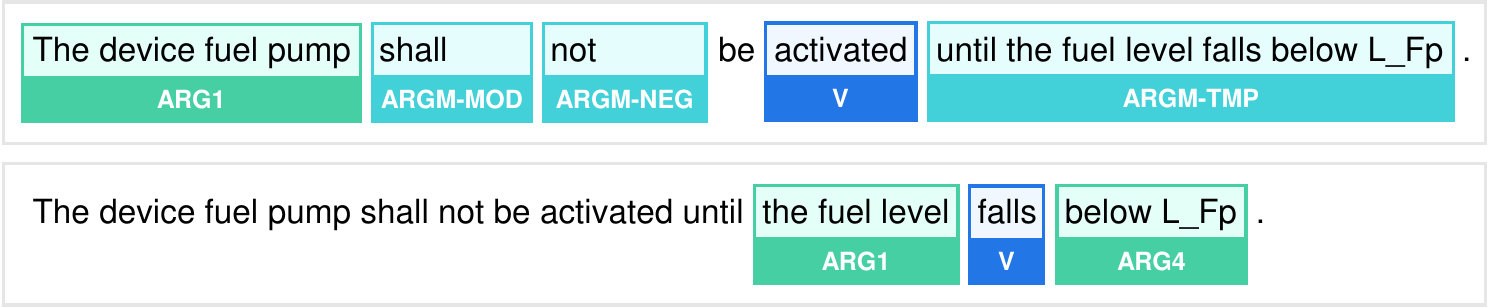}
\caption{SRL frames for Req.\,\ref{req:bomb_2}.}
\label{fig:srl_frames2}
\end{figure}

Consider another requirement, Req.~\ref{req:bomb_2}, with negation and the conditional keyword \textit{``until"}. SRL frames obtained from the pre-trained model are shown in Fig.~\ref{fig:srl_frames2}. \vspace{10pt}

\paragraph{Entity detection (negation handling)}
The first frame follows the same construct as before leaving the argument \textit{ARG2} optional. However, we also have the argument \textit{ARGM-NEG} indicating a negation in the clause which should be considered while forming the relation. In this case, we can modify the previous construct \verb|V_ARG1()| to form a new construct \verb|not V_ARG1()|. This finally yields the relation \verb|not activate_device_fuel_pump()|.

Though the second frame indicates a conditional clause, it is difficult to identify since no arguments have been assigned to the word \textit{``until''}. So as to recognize it as a condition, we apply rules to identify it as condition like keyword identification. Considering the arguments, we have the role \textit{ARG4}, which indicates an ending point that can be mapped to a parameter and \textit{ARG1} to a signal. \vspace{10pt}

\paragraph{Operator detection}
To find the operator symbol, the verb text will not be enough in this case as \textit{``falls"} alone cannot help in getting the operator symbol. So, as to get the symbol, we apply one extra rule, i.e., to consider the next word in the requirement in addition to the verb text to get the symbol. So the span text would be \textit{``falls below"} which gives the symbol ``$<$''.

So this particular frame would lead to \verb|fuel_level() < L_Fp| leading to the formation of the construct \verb|ARG1 V ARG4|.

\begin{requirement} \label{req:bomb_3}
The device fuel pump shall be deactivated within 3s and shall be closed when the fuel level exceeds [L\_Fp].
\end{requirement}

\begin{figure}
\centering
\includegraphics[width=\linewidth]{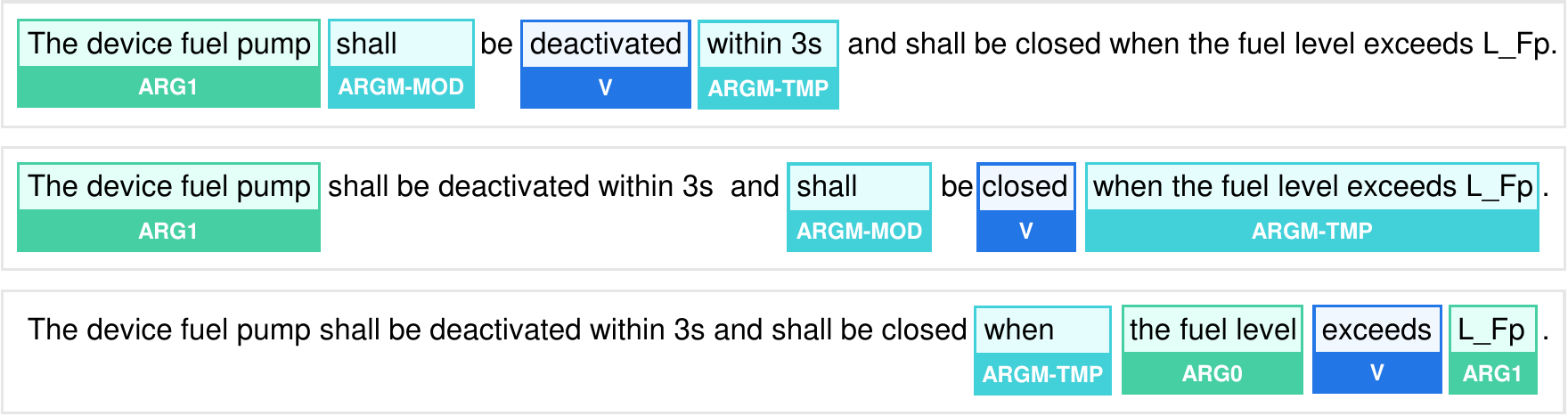}
\caption{SRL frames for Req.\,\ref{req:bomb_3}.}
\label{fig:srl_frames3}
\end{figure}

SRL can handle the time constraint and some complex decomposition as well. To demonstrate this, we will modify the previous requirement as shown in Req.\,\ref{req:bomb_3}. Figure \ref{fig:srl_frames3} shows the identified SRL frames. \vspace{10pt}

\paragraph{Decomposition}
When we decompose this requirement with keywords like \textit{``and"} and \textit{``when"}, it would lead to wrong formation of clause as the second clause would not show what shall be closed. Looking at the SRL frames, it correctly identifies this and forms the correct following clause by considering span text for each frame. 
\begin{itemize}
    \item The device fuel pump shall deactivated within 3s
    \item The device fuel pump shall closed
    \item when the fuel level exceeds L\_Fp
\end{itemize}

The first and the second frame follows the same construct as before, i.e., \verb|V_ARG1()|, and the third frame follows \verb|ARG0 V ARG4|. In the first frame, we have identified \textit{ARGM-TMP} with a time constraint. This temporal behavior can be detected by some rule, e.g. whether the span text contains a number and a unit of time, or from the keyword \textit{``within''}. However, we leave the discussion of modeling non-functional properties such as temporal behavior to future work.



In Table \ref{tab:srl_constructs} we have summarized some rules for mapping the arguments to model entities. The first three lines in the table show some cases for assignments, whereas the two last lines show cases for conditions. Conditions are identified using the argument \textit{ARGM-TMP}.
The \textit{operator} is identified from the \textit{V} role denoted by op(V) in the table.

\begin{table}
\caption{Mapping of SRL arguments to model entities}
\label{tab:srl_constructs}
\centering
\begin{tabular}{ l@{\extracolsep{10pt}}l@{\extracolsep{20pt}}l@{\extracolsep{10pt}}l}
\toprule
\textbf{SRL arguments} & \textbf{Signal} & \textbf{Operator} & \textbf{Parameter} \\
\midrule
 ARG1 (patient), V & V\_ARG1 & - & - \\
ARG1 (patient), V, ARG2 (instrument) & V\_ARG1 & - & ARG2 \\
 ARG1 (patient), V, ARGM-PRD (state) & V\_ARG1 & - & ARGM-PRD \\
\midrule
ARG0 (agent), V, ARG1 (patient) & ARG0 & op(V) & ARG1 \\
ARG1 (agent), V, ARG4 (end points) & ARG1 & op(V) & ARG4 \\
\bottomrule
\end{tabular}
\end{table}

\paragraph{Discussion}
Using SRL information, we need much easier and less rules as compared to the first approach. This makes it easy to adapt this approach according to a specific use case and needs. However, in a few cases we found that SRL was not working properly. For further improvements, one could of course combine both approaches using the POS and dependency tags as well as SRL labels.

Similar as in the first approach, the underlying deep learning models are evolving quickly. They are trained with new and better annotated datasets, resulting in better generation of models with state-of-the-art performance and higher accuracy. The above mentioned constructs and rules are based on the specific model (current state of the art \cite{shi2019simple}) we used to extract SRL arguments. This might not work completely with the future better models as they would extract more information and these constructs/rules could be outdated. So it might be necessary to change the rules when other models are used.

\subsection{Model Formation}

Once the necessary entities are extracted from the requirement clauses, we head to the last stage in the pipeline, model formation.
As mentioned above, we form a relation for each clause. In particular, assignments are of the form \verb|signal(parameter*)| with optional parameter, and conditional statements are of the form \verb|signal()| \verb|operator| \verb|parameter|.
To generate requirement models from these relations, we use a simple domain-specific language (DSL) with abstract logical blocks. We extend the modalities of our previous work \cite{gropler2022automated} following the Temporal Action Language (TeAL) of Li et al.\ \cite{li2015towards}.
It maps the relations according to some identified keywords to blocks. Conditional statements starting with \textit{if}/\textit{when}/\textit{while} are mapped to an \texttt{if}-block, those starting with \textit{until} to an \texttt{until}-block.
Assignments without introductory keyword are mapped to an \texttt{then}-block and those starting with \textit{else}/\textit{otherwise} to an \texttt{else}-block.
When there is no conditional statement given in the requirement, we map all assignments to a statement-block without using any keywords.
Conjunctions (\textit{and}/\textit{or}) identified between relations are also accommodated in these DSL blocks. Though this mapping appears rather trivial, our aim is to make this translation simple and flexible, so that ways are open for integration with other sophisticated languages. Obviously, our DSL is very similar to the FRETISH language \cite{giannakopoulou2020generation}. The resulting models can also be further transformed into Matlab Simulink models or UML sequence diagrams, depending on what the end user desires. 
 


For example, the model for Req. \ref{req:akka_1} is of the form:
\begin{lstlisting}
    if( temperature_of_battery() > t_batt_max )
        then( set_error_state( E_BROKEN) )
\end{lstlisting}
Similarly, the more complex requirement defined as Req. \ref{req:bomb_1} will yield the following model:
\begin{lstlisting}
    if( device_temperature() > T_Hi )
        then( limit_maximum_power(G_Max)
            and indicate_event_high_device_temperature() )
\end{lstlisting}

\begin{requirement} \label{req:bomb_4}
The maximum power shall be limited to [G\_Max] when the device temperature exceeds [T\_Hi] ºC, otherwise indicate the error ``Maximum power exceeded", until the device temperature falls below [T\_Norm] ºC.
\end{requirement}

To demonstrate how we build the DSL with all the logical blocks, see Req. \ref{req:bomb_4}. This requirement has an \texttt{if}-block, its corresponding \texttt{then}-block, an \texttt{else}-block and an \texttt{until}-block. The formal model for this requirement is of the following form:
\begin{lstlisting}
if( device_temperature() > T_Hi )
    then( limit_maximum_power(G_Max) )
else( indicate_error_maximum_power_exceeded() )
until( device_temperature() < T_Norm )
\end{lstlisting}








\paragraph{Discussion}
There exist many different, more or less common textual and graphical representations of behavioral models, for example temporal logic, cause-effect graphs, UML diagrams (sequence diagrams, class diagrams, activity diagrams, etc.), OCL constraints, Petri nets, or Matlab Simulink models.
Initially, we planned to use sequence diagrams as output format since it is very common and can be represented textually and graphically\footnote{\url{https://sequencediagram.org/}}. However, we realized that some parts of the requirements, such as \textit{until}-conditions or temporal behavior, cannot be handled easily. For more complex requirements, the overview gets lost due to many nested fragments. 
Pseudocode has the advantage of being human- and machine-readable.
It is a kind of more abstract, intermediate format, one has to define a mapping for the keywords for further processing.
We see this as an advantage, since this mapping can be adapted to different use cases and needs. It includes all information to be mapped to other model formats.

\section{Existing Methodologies}
\label{sec:related_work}

In this section, we want to give a brief overview of existing methodologies for requirements formalization.

\subsection{Controlled Natural Language (CNL)}
The use of a Controlled Natural Language (CNL) is an old and common approach to enable automatic analysis. The requirements need to be written in restricted vocabulary and following some specific templates (also called patterns or boilerplates). Commonly used templates include the SOPHIST template \cite{rupp2014requirements}, the EARS template \cite{mavin2009easy} and the FRETISH template \cite{giannakopoulou2020generation}. 


Early NLP methods used dictionaries and a large set of rules to extract syntactic and semantic information such as syntax trees and semantic roles to form models \cite{riebisch2005traceability,carvalho2015nat2test}. More recently, Allala et al.\ \cite{allala2019towards} exploit templates to represent requirements and utilize Stanford CoreNLP to create test cases and meta-modeling for validation. 
The FRET tool can be used to map the structured requirement with the help of templates and convert them into formal models \cite{farrell2022fretting}. Giannakopoulou et al.\ \cite{giannakopoulou2020generation} use FRETISH requirements and derive temporal logic formulas from them. However, these approaches use rule sets rather than any kind of NLP method.

\subsection{Dependency and Part of Speech (POS)}

As explained in Section \ref{sec:dep-pos}, part-of-speech (POS) tags categorize words according to their grammatical properties, and dependency tags identify grammatical structure of the sentence. Different libraries like NLTK, spaCy and Stanford CoreNLP can be used for POS tagging and dependency parsing. Dependency annotation can also be done using Universal Dependencies Guidelines on software requirements \cite{hassert2021ud}. 

Pre-defined patterns are considered on dependency trees to identify subtrees which would help in generation of cause-effect-graphs \cite{fischbach2020specmate}.
Unstructured Information Management Architecture (UIMA) framework has been used to extract POS tags and determine phrases to help identify actions and conditions to form models on use case descriptions \cite{sinha2009linguistic}.
Fischbach et al.\ \cite{fischbach2020specmate}
also constructs few patterns with combination of POS and dependency tags.
Koscinski et al.\ \cite{koscinski2021natural} use Stanford CoreNLP to assign dependency tags and use pre-defined patterns to identify syntactical categories from which requirements models can be generated with a small amount of rules. 
Fritz et al.\ \cite{fritz2021automatic} uses the combination of both POS and dependency tags  to identify semantic frames to map requirements into an existing structure. 
The latter two papers are very similar to our first approach, but use different categories and rules.

\subsection{Named Entity Recognition (NER)}

Phrases from sentences can be classified into some predefined categories known as named entities. NER is the task of assigning these entities. Commonly used categories are person, organization, location etc. But these categories are not helpful for requirements analysis. To perform NER for requirement texts, one needs to define new categories and manually annotate a dataset to train ML-based approaches. Annotating a dataset is a laborious and a time consuming task. Also, defining categories can vary depending on the requirement type. Malik et al.\ \cite{malik2021named} and Herwanto et al.\ \cite{herwanto2021named} defined 10 and 5 entities, respectively, for software requirement specifications and privacy related tasks. In both approaches, the dataset is annotated manually. Multiple checks with multiple human annotators are carried out to get better quality data. Nayak et al.\ \cite{nayak2022req2spec} have trained an NER model for creating expressions with 9 entity categories, which is then used to formalize requirements.





\subsection{Semantic Role Labeling (SRL)}

As described in Section \ref{sec:srl-based}, in Semantic Role Labeling (SRL), words or phrases in a sentence are assigned labels that indicate their semantic role in the sentence, such as the agent, the patient, etc.
Semantic frames have been widely used for requirements analysis in recent years. From simple dependency parsing rules \cite{sengupta2015verb} to machine learning models \cite{bjorkelund2009multilingual,shi2019simple}, much has been used to extract semantic roles and frames.
Sengupta et al.\ \cite{sengupta2015verb} use the Stanford Dependency manual to create rules that help extract basic arguments for requirement analysis, whereas Fritz et al.\ \cite{fritz2021automatic} use spaCy to extract POS and dependency tags from which they create rules to extract semantic roles. Rules can also be created to recognize semantic relations based on the lexeme (word form) and its variants \cite{bokaeihosseini2018inferring}.

FrameNet was utilized to determine semantic relationships between requirements with inheritance knowledge \cite{alhoshan2018using}.
VerbNet and other open source resources were used to perform semantic analysis for specification representation \cite{mahmud2017specification}.

Recently, machine learning techniques have also been considered for semantic role extraction. Wang et al.\ \cite{wang2022automatic} use the SRL model of the CogComp NLP pipeline to generate domain models with a rule set. Mate Tools \cite{bjorkelund2009multilingual} has been a commonly used ML model for SRL which uses linear logistic regression.
Diamantopoulos et al.\ \cite{diamantopoulos2017software} extended Mate tools' semantic role labeler by training it with additional lexical attributes as features, and used it for mapping of requirements to formal representations. The latter work is similar to our second approach but we utilize a much more recent SRL model with very flexible rules.

\subsection{Translation and Back-Translation}

Translation of requirement texts to some kind of model directly in one step is not an easy task. As seen from all the above mentioned methods, it requires multiple steps and multiple methods to form the output model. We can use the concept of Neural Machine Translation (NMT)/Code Generation to translate requirements to desired pseudocode. Code generation has been used to convert complex text into programming language code using machine learning models \cite{ling2016latent}. It can also be used to convert text into an Abstract Syntax Tree (AST) which represents syntactic structure of the desired code \cite{rabinovich2017abstract}. This would require manual creation of a large dataset with requirement texts and its corresponding pseudocode or AST which is again a laborious and a time consuming task.

Neural Machine Translation can also be used to back translate output models to requirement text. Tools like OpenNMT has been used to convert temporal logic into natural language requirements \cite{cherukuri2022towards}. A set of rules can also be applied to create functional natural language requirements from data models \cite{brock2022nlg4re}. A systematic survey on the generation of textual requirements specifications from models can be found in \cite{nicolas2009on}.

\section{Discussion} 
\label{sec:discussion}

In view of the many recent developments in this area of research, the journey towards the development of mature requirements formalization techniques has only just begun. Since the aim of this work is to present principal ideas on this research topic, we also want to discuss some findings and shortcomings that we identified during our investigations.
The tremendous survey of Zhao et al.\ \cite{zhao2021natural} has already identified some key findings for the NLP-supported requirements engineering process, namely (i) a huge gap between the state of the art and the state of the practice, i.e., an insufficient industrial evaluation, (ii) little evidence of industrial adoption of the proposed tools, (iii) the lack of shared RE-specific language resources, and (iv) the lack of NLP expertise in RE research to advise on the choice of NLP technologies.
We agree that these findings also apply to the requirements formalization process.

\subsection{Lessons Learned}
The international XIVT project\footnote{\url{https://www.xivt.org/}} addressed the challenge of automatically analyzing and extracting requirements in order to minimize the time for testing highly configurable, variant-rich systems. 
During the three-year project period, we had many interesting talks and discussions with research and industry partners on this research area.
The approaches presented in this work are the result of countless internal and external discussions, which shows that it is not an easy task and very manifold, but also a very interesting topic. \vspace{10pt}

\paragraph{Generality}
It is very tempting to try to develop a single algorithm that is general enough to handle a wide range of natural language requirements. But, just as natural language is diverse, so would the algorithm to be developed have to be. Requirements are formulated by such a wide variety of people with different motivations, experience and expert knowledge, from different branches and with different social backgrounds, that it is currently impossible to create one algorithm for the huge variety of styles of unstructured natural language.
The many existing approaches and our own developments show that it is reasonable to automate this process to a high degree for a specific use case, domain and writing style.
In short: \textit{Don't worry about being specific!} \vspace{0pt}

\paragraph{Requirement quality}
The human language is full of misunderstandings, ambiguities, inconsistent and incomplete formulations. When developing a rule set we were strongly biased by the given requirements. We kept trying to improve the rules to make them more and more accurate. 
In discussion with our industry partners, we realized that some parts of the requirements seemed to be somewhat poorly written. By a short reformulation, our algorithm could handle them much easier. This shows that the algorithmic treatment of the formalization process reveals many minor issues that need to be fixed by the requirement engineers, and thus also contributes to a higher quality of the requirements.
In short: \textit{Don't always blame the model!} \vspace{10pt}

\paragraph{Degree of automation} 
For a long time of our development, we planned to build a standard input/output tool that would generate the requirements models fully automatically without asking the user for any input or confirmation. However, we have found that we are unable to properly process the data if some information within the pipeline is missing. 
Additionally, our industry partners have also requested to always have a quick verification option by prompting missing or uncertain information to the user.
So if even humans need to review many details to formalize a set of requirements, how should an algorithm be able to generate correct and complete models fully automatically?
The fact that requirements analysis is a highly interactive process should also be taken into account when implementing the tool.
It is much more desirable from the industry to have assistance in creating correct requirements models than to have a fully automated algorithm that produces some erroneous models.
In the end, the required working time of an engineer decides whether the review of automatically generated models is preferable to the manual creation process.
That is, the algorithms need to be highly efficient to have the chance of being used by the industry, therefore we believe that an interactive, performant user interface is essential.
In addition, the information obtained from user input can also be used to improve the underlying ML algorithms.
In short:  \textit{Go for user interaction!} \vspace{10pt} 

\paragraph{Availability of data}
A crucial obstacle in this research area is that the data situation for freely available requirements with corresponding models is very poor. One would need a large corpus of these data from different use cases to compare different algorithms. Unless a reliable and sufficiently large amount of ground truth exists, a benchmark of different approaches and tools is not possible \cite{tichy2015nlrpbench}.
Even if one takes the time in a group of scientists to analyze, label, and model requirements, detailed discussions with industry stakeholders are essential and time-consuming.
Consider the quote of Steve Jobs: \textit{"A lot of times, people don't know what they want until you show it to them"}\footnote{Business Week 12 May 1998, \url{https://www.oxfordreference.com/view/10.1093/acref/9780191826719.001.0001/q-oro-ed4-00005922}}.
Thus, the success of ML-trained algorithms relies on having a large amount of labeled and unlabeled data sets available.
In short: \textit{Share your data!} \vspace{0pt} 

\paragraph{Availability of tools}
Similarly, the availability of ready-to-use algorithms and tools is very limited.
Only a few tools are publicly available \cite{zhao2021natural}, and most of them are in a prototypical stage.
This makes it very difficult for industry to identify useful tools for the application to their use cases.
In short: \textit{Go for GitHub!} \vspace{10pt} 

\paragraph{Evolution}
Most approaches in the literature deal with a fixed set of requirements. Likewise, most freely available requirements specifications are provided in a single fixed document (with a few exceptions\footnote{AUTOSAR, \url{https://www.autosar.org/nc/document-search/}}). However, this is completely contrary to the highly iterative and evolving process of requirements elicitation. 
Just as dynamic as software development is the corresponding requirements engineering.
On the other hand, we observe that NLP models have evolved very rapidly in recent years. Since the publication of the BERT models, there has been an exploding amount of research about the application and further training of these models. Researchers are hunting for better and better models with higher accuracy, algorithms from one or two years ago are already outdated and need to be revised when used for further development.
In short: \textit{Everything evolves!}

\subsection{Possible Future Developments}
From our perspective, there are a number of emerging, sophisticated NLP methods and models that can be applied to the requirements formalization process. 
Although it is very difficult to look into the distant future of NLP research and predict developments, we do see some direct follow up to current approaches and can envision some future developments. \vspace{6pt}

\paragraph{Named Entity Recognition (NER)}
The advantage of NER is the direct identification of model entities when they are used as labels. However, the labeled dataset must be very large to achieve reasonable accuracy. To reduce the intensive labelling work, one could use pre-trained models (e.g.\ provided by AllenNLP) and fine-tune them with newly annotated datasets with new categories. \vspace{6pt}

\paragraph{Semantic Role Labeling (SRL)}
The current SRL models already have an acceptable accuracy. However, we also found that the predictions can change suddenly if just one word or comma is changed. The model may not work as well for domain-specific wording or the writing style of requirements. Therefore, the improvement in accuracy could be studied if the SRL models are pre-trained with some labeled requirements data (a starting point could be \cite{diamantopoulos2017software}. \vspace{6pt}

\paragraph{Question-Answer Driven Semantic Role Labeling (QA-SRL)}
As mentioned above, requirements analysis is highly interactive and needs several input from the user. It would be a new level of automation, if an algorithm itself is able to formulate questions and process the given answers to clarify missing or uncertain information for generating the model. \vspace{6pt}

\paragraph{Transfer learning (TL)}
It is very successful in other fields of NLP research to train a model an some related tasks with a larger amount of data and then to fine-tune it for the actual task with much less labeled data. We could imagine using models that have already been trained for requirements classification or ambiguity detection. \vspace{6pt}

\paragraph{Online learning}
Current approaches use NLP models that are trained once and will not change according to user feedback. It would be much more convenient for the user if the algorithm learns from the input and suggests better results for subsequent requirements. This could be done by some sort of post-training of the models or more sophisticated continuous learning methods. This could also be useful for a completely new use case and domain, where the algorithm learns the rules from the beginning and one could avoid implementing of a rule set at all. \vspace{6pt}

\paragraph{Translation/back-translation}
Directly translating requirements into models with a large ML model might be far in the future or somehow not reasonable. But there are interesting developments in the research area of translating text to code that could be helpful in formalizing requirements as well. The other direction of generating boilerplates sounds much simpler, but also needs to be studied in detail. We can imagine that in the future a coexistence of requirements and models will be possible, where hand-written requirements are transformed into models and transformed back into structured text that can be used for further processing - a kind of "speaking models". \vspace{6pt}
    
\paragraph{Semantic similarity}
As shown in our previous work \cite{gropler2022automated}, semantic similarity between requirements and product design descriptions (if available) is helpful for generating more concrete requirements models, i.e., identifying signal and parameter names from the design specification instead of generating abstract entity names. However, identifying the correct Boolean value (antonym detection), for example, is still a difficult task in NLP research. Moreover, requirements formalization and quality analysis are strongly interlinked. Semantic similarity can be useful for consistency checking, e.g.\ for identifying of identical signals and parameters used with slightly different terms within the set of requirements, see \cite{wang2022automatic} for a first approach. \vspace{6pt}

\paragraph{Auxiliary documents}
Requirements do not fall from the sky. Usually, an extensive elicitation phase takes place in advance. One could make use of preparatory documents such as interviews, meeting notes, etc., or implementation artifacts such as issue descriptions (bug reports, feature requests, etc.). They can provide context, different formulations, and missing information.
It would also be interesting to explore how to support the entire requirements development process. One would need to study the process in great detail, track and analyze the documents, and identify some successful workflows that can be assisted by NLP approaches (a starting point could be the FRET tool \cite{giannakopoulou2020generation}).

\section{Conclusion} 
\label{sec:conclusion}

In this work, we have presented and discussed principal ideas and state-of-the-art methodologies from the field of NLP to automatically generate requirements models from natural language requirements.
We have presented an NLP pipeline and demonstrated the iterative development of rules based on NLP information in order to guide readers on how to create their own set of rules and methods according to their specific use cases and needs. 
We have studied two different approaches in detail. The first approach, using dependency and POS tags, shows good results but is somewhat inflexible in adopting the underlying rule set for different use cases and domains. The second approach, using semantic roles, shows similar good results, but requires less effort to create a set of rules and can be easily adapted to specific use cases and domains.
The use of a human- and machine-readable format for the requirements models appears suitable for easy reading comprehension and at the same time for automatic further processing.
The results show that the current pre-trained NLP models are suitable to automate the requirements formalization process to a high degree. 

Furthermore, we provided an overview of the literature and recent developments in this area of research and discussed some findings and shortcomings (lessons learned) that we identified throughout the development of our approaches. Finally, we suggested some possible future developments in NLP research for an improved requirements formalization process.


\footnotesize{
\paragraph{Acknowledgments}
This research was funded by the German Federal Ministry of Education and Research (BMBF) within the ITEA projects XIVT (grant no.~01IS18059E) and SmartDelta (grant no.~01IS21083E). We thank AKKA Germany GmbH and Alstom for providing industrial use cases for the demonstration of the presented methods.
We are grateful to Prof.~Holger Schlingloff from Fraunhofer FOKUS/HU Berlin for his valuable recommendations.}

\bibliographystyle{splncs04}
\bibliography{NLP_references}

\end{document}